\documentclass{article}

\usepackage{amssymb}
\usepackage{verbatim}
\usepackage[francais,english]{babel}
\usepackage{graphicx}
\usepackage{graphics}
\usepackage{lineno}
\usepackage[latin1]{inputenc}
\usepackage[table]{xcolor}
\usepackage[]{natbib}
\usepackage[]{multirow}
\usepackage[]{a4wide}
\usepackage{fancyhdr}
\usepackage{lastpage}

\begin{document}

\baselineskip 0.8cm

\title{Fast Inference of Admixture Coefficients Using Sparse Non-negative Matrix Factorization Algorithms}

\date{}

\author{Eric Frichot$^1$, Fran\c cois Mathieu$^1$,  Th\'eo Trouillon$^{1,2}$,\\
 Guillaume Bouchard$^2$, Olivier Fran\c cois$^1$}

\maketitle

\begin{center}
$^1$ Universit\'e Joseph Fourier Grenoble 1, Centre National de la Recherche Scientifique, TIMC-IMAG UMR 5525, 38042 Grenoble, France.\\
$^2$ Xerox Research Center Europe, F38240 Meylan, France. 
\end{center}

{\noindent Corresponding Author:\\ Olivier Fran\c cois \\
    Faculty of Medicine \\
    Grenoble\\
    F38706 La Tronche\\
    France \\
    +334 56 52 00 25 (ph.)  \\
    +334 56 52 00 55 (fax)  \\
    \texttt{olivier.francois@imag.fr}}
{\\Running Title: Super fast inference of population structure\\}

\clearpage
\newpage 
\begin{center}
Abstract
\end{center}

Inference of individual admixture coefficients, which is important for population genetic and association studies, is commonly performed using compute-intensive likelihood algorithms. With the availability of large population genomic data sets, fast versions of likelihood algorithms have attracted considerable attention. Reducing the computational burden of estimation algorithms remains, however, a major challenge. Here, we present a fast and efficient method for estimating individual admixture coefficients based on sparse non-negative matrix factorization algorithms. We implemented our method in the computer program {\tt sNMF}, and applied it to human and plant genomic data sets. The performances of {\tt sNMF} were then compared to the likelihood algorithm implemented in the computer program {\tt ADMIXTURE}.  Without loss of accuracy, {\tt sNMF} computed estimates of admixture coefficients within run-times approximately 10 to 30 times faster than those of {\tt ADMIXTURE}.

\newpage 

\section{Introduction}

Inference of population structure from multi-locus genotype data is commonly performed using likelihood methods implemented in the computer programs {\tt STRUCTURE}, {\tt FRAPPE} and {\tt ADMIXTURE} (Pritchard et al. 2000a, Tang et al. 2005, Alexander et al. 2009). These programs compute probabilistic quantities called {\it admixture coefficients} that represent the proportions of an individual genome that originate from multiple ancestral gene pools. Estimation of admixture proportions is important with many respects, for example in delineating genetic clusters, drawing inference about the history of a species, screening genomes for signatures of natural selection, and performing statistical corrections in genome-wide association studies (Pritchard et al. 2000b; Marchini et al. 2004; Price et al. 2006; Frichot et al. 2013).

Individual admixture coefficients can be estimated using either supervised or unsupervised statistical methods. Supervised estimation methods use predefined source populations as ancestral populations. Classical supervised estimation approaches were based on least-square regression of allele frequencies in hybrid and source populations (Roberts and Hiorns 1965; Cavalli-Sforza and Bodmer 1971). Unsupervised approaches attempt to infer ancestral gene pools from the data using likelihood methods. An undesired feature of likelihood methods is that they can be computer intensive, with typical runs lasting several hours or more. With the use of dense genomic data and increased sample sizes, reducing the time lag necessary to perform estimation is a major challenge in population genetic data analysis. 

A fast approach to the estimation of ancestry coefficients is by using principal component analysis (PCA, Patterson et al. 2006). PCA is an exploratory method that describes high-dimensional data using a small number of dimensions, and makes no assumptions about sampled and ancestral populations. Using PCA can lead to results surprisingly close to likelihood methods, and connections between methods have been intensively investigated during the last years (Patterson et al. 2006; Engelhardt and Stephens 2010; Frichot et al. 2012; Lawson et al. 2012; Lawson and Falush 2012). But a drawback of PCA is that interpretation in terms of ancestry is often difficult, as it can be confounded by demographic factors or irregular sampling designs (Novembre and Stephens 2008; Mc Vean 2009; Fran\c cois et al. 2010).

In this study, we introduce computationally fast algorithms that lead to estimates of admixture proportions comparable to those obtained with {\tt STRUCTURE} or {\tt ADMIXTURE}. The algorithms were implemented in the computer program {\tt sNMF} based on  sparse non-negative matrix factorization (NMF) and least-square optimization (Lee and Seung 1999; Kim and Park 2007; Kim and Park 2011). Like PCA, sparse NMF algorithms are flexible approaches that are robust to departures from traditional population genetic model assumptions. In addition, sparse NMF algorithms produce estimates of admixture proportion within run-times that are much faster than {\tt STRUCTURE} or {\tt ADMIXTURE}. This study assesses the utility of sparse NMF algorithms when analyzing population genetic data sets, and compares the performances of {\tt sNMF} algorithms with those implemented in {\tt ADMIXTURE} on the basis of human and plant data.

\clearpage
\newpage

\section{Results}

\begin{table}
\caption{{\bf Data sets used in this study.}}
\begin{center}
\begin{tabular}{lccc}
\hline
{\bf Data set} &   Sample size & Number of SNPs &  Reference \\
\hline
 &  &   & \\
  HGDP00778 & 934 & 78K  & (Patterson et al. 2012) \\
 HGDP00542 &  934 & 48.5K & -- \\
 HGDP00927 & 934  & 124K & --\\
 HGDP00998 & 934 & 2.6K& -- \\
 HGDP01224 & 934 & 10.6K & -- \\
 HGDP-CEPH & 1,043 & 660K & (Li et al. 2008)\\
 1000 Genomes  & 1,092 & 2.2M & (The 1000 Genomes Project Consortium 2012)\\
 {\it A. thaliana} &  168  & 216K & (Atwell et al. 2010)\\
  &  &   & \\
 \hline
\end{tabular}
\end{center}
\end{table}

We used the program {\tt sNMF} to implement sparse non-negative matrix factorization and to compute least-squares estimates of ancestry coefficients for world-wide human population samples and for European populations of the plant species {\it Arabidopsis thaliana}. As in the likelihood model implemented in the computer programs {\tt STRUCTURE} and {\tt ADMIXTURE}, {\tt sNMF} supposes that the genetic data originate from the admixture of $K$ parental populations, where $K$ is unknown, and it returns estimates of admixture proportions for each multi-locus genotype in the sample (Pritchard et al. 2000a, Alexander et al. 2009). To perform the estimation of admixture coefficients, the {\tt sNMF} algorithm solves penalized versions of a least squares minimization problem using an active set method (Kim and Park 2011, see Materials and Methods).

\paragraph{Accuracy of admixture estimates on HGPD data.}  First we evaluated the ability of {\tt ADMIXTURE} estimates to be accurately reproduced by {\tt sNMF} algorithms for $5$ Harvard HGDP panels and for the HGDP-CEPH data set (Li et al. 2008; Patterson et al. 2012).  For each run of {\tt ADMIXTURE}, we computed a maximum squared correlation coefficient ($R^2$) and a minimum root mean squared error (RMSE) over runs of {\tt sNMF} performed with identical number of clusters ($K$). For $K$ ranging from 5 to 10, squared correlation coefficients remained greater than 0.96 on average across all runs (Figure 1). Average values of the RMSE remained lower than $5.5\%$ across all runs (480 runs, Table S1). These results provided evidence that {\tt sNMF} estimates closely reproduce those obtained with {\tt ADMIXTURE} across the 6 HGDP data sets.  

\includegraphics{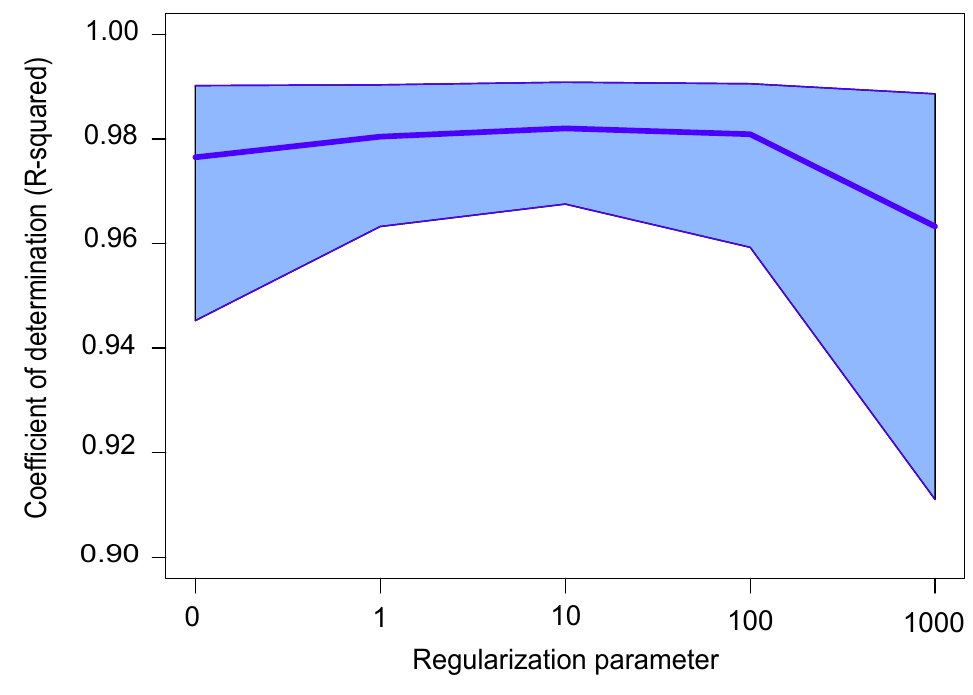}
\\
{\it 
\noindent{\bf Figure 1.} {\bf Correlation between {\tt sNMF} and {\tt ADMIXTURE} estimates.} Squared correlation (coefficient of determination, $R^2$) between the admixture coefficients estimated by each program. For each number of clusters ($K$), the result corresponds to the maximum correlation over 5 runs, averaged over values of the regularization parameter lower than 1,000 and over 6 HGDP data sets. The shaded area corresponds to a $95\%$ confidence interval displayed for each value of the regularization parameter, $\alpha$.
}

\paragraph{Run-time analysis.} Next we performed run-time analyses for {\tt ADMIXTURE} and for {\tt sNMF} using the previous HGDP data sets in addition to The 1000 Genomes Project phase 1 data. When running programs, the default stopping criterion of each algorithm was implemented. The averaged run-times were averaged over distinct random seed values for each value of $K$. Run-times increased with the number of SNPs in the data set and with the number of clusters in each algorithm (Figure 2, Table 2, Figure S1). For data set HGDP01224 (10.6K SNPs), it took on average 0.8 minute (1.7 minutes) to {\tt sNMF} to compute admixture estimates for $K = 10$ ($K = 20$) clusters. For {\tt ADMIXTURE}, the run-time was on average equal to 11 minutes (48 minutes) for $K = 10$ ($K = 20$) clusters. For panel HGDP00778 (78K SNPs), it took on average 7.2 minutes (15 minutes) to {\tt sNMF} to compute admixture estimates for $K = 10$ ($K = 20$) clusters. For {\tt ADMIXTURE}, the average run-time was 1.5 hour (6.2 hours) for $K = 10$ ($K = 20$) clusters. For the CEPH-HGDP data sets (660K SNPs), it took on average 55 minutes (2.1 hours) to {\tt sNMF} to compute ancestry estimates for $K = 10$ ($K = 20$) clusters. For {\tt ADMIXTURE}, the average run-time was 12 hours (38 hours) for $K = 10$ ($K = 20$) clusters. Run-times increased in a quadratic fashion with $K$ in {\tt ADMIXTURE} whereas they increased linearly in {\tt sNMF} (Figure 2). For the values of $K$ used in our analyses, {\tt sNMF} ran 5 to 30 times faster than {\tt ADMIXTURE} when these programs were applied to HGPD data sets. Regarding The 1000 Genomes Project phase 1 data set, the average run-times of {\tt sNMF} were approximately equal to 2.8 hours (4.6 hours) for number of clusters $K = 5$  $(K = 10)$. The {\tt ADMIXTURE} runs led to similar estimates of $Q$, but a single run on the phase 1 data set took more than 19 hours for $K = 5$ (59 hours for $K = 10$). 

\centerline{\includegraphics{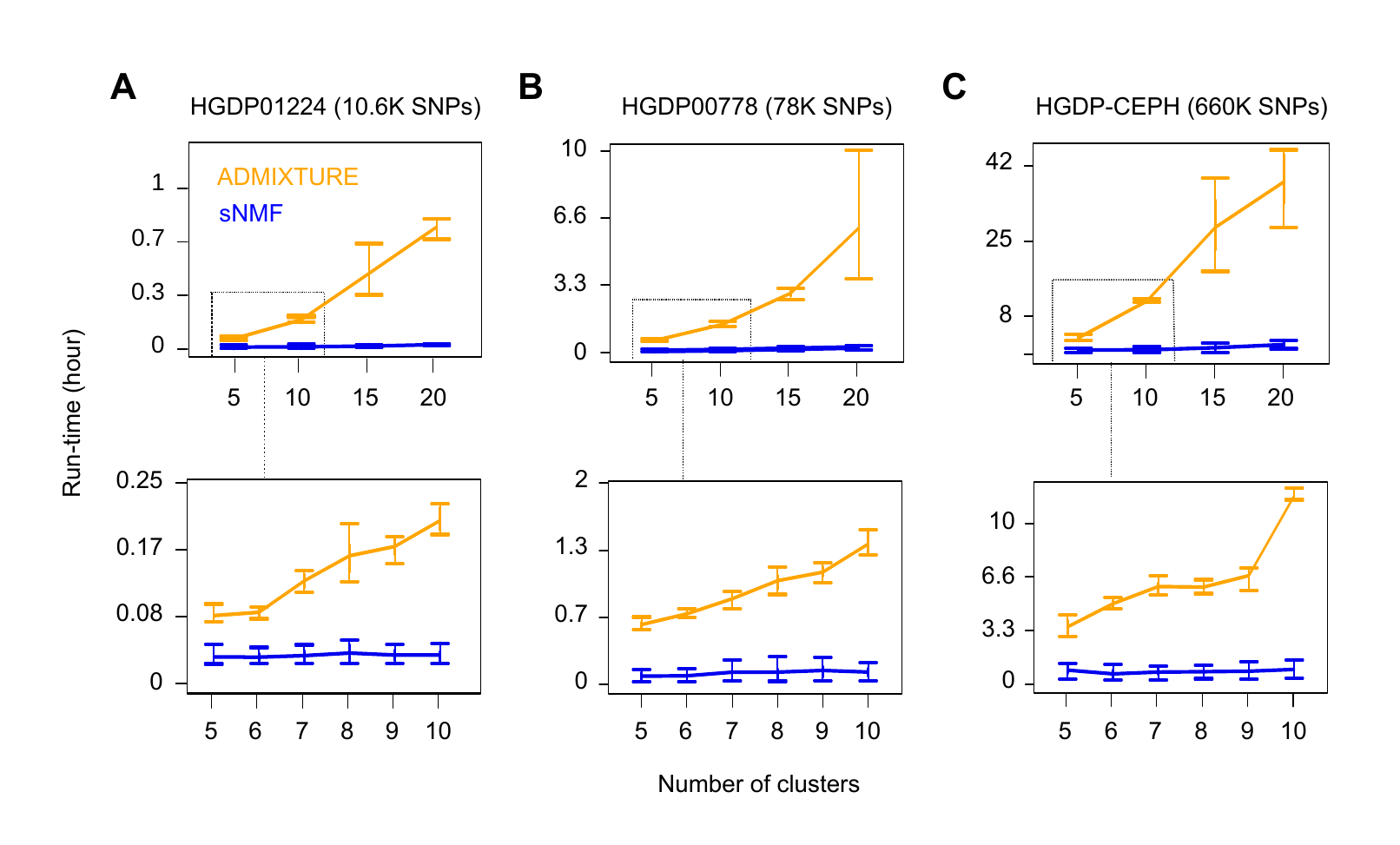}}
{\it 
\noindent{\bf Figure 2.} {\bf Run-times for {\tt sNMF} and {\tt ADMIXTURE} runs.} Averaged time elapsed before the stopping criterion of the {\tt sNMF} (blue) and {\tt ADMIXTURE} (orange) programs is met. Time is expressed in unit of hours. A) Run-time analysis for Harvard HGDP panel 01224 (10.6K SNPs). B) Run-time analysis for Harvard HGDP panel 00778 (78K SNPs). C) Run-time analysis for the HGDP-CEPH data (660K SNPs).
}

\begin{table}
\caption{{\bf   Run-time summary for sNMF and ADMIXTURE.} Average values and their 95\% confidence intervals.
}
\vspace{0.8cm}
\hspace{-1.5cm}
\begin{tabular}{cccccccc}
\hline
{\bf Data set}  & \footnotesize time & \multicolumn{2}{c}{$K=5$} & \multicolumn{2}{c}{$K=10$} &  \multicolumn{2}{c}{$K=20$} \\
(Number of SNPs)& \footnotesize unit & sNMF  & ADMIXTURE & sNMF & ADMIXTURE & sNMF & ADMIXTURE\\
\hline

 HGDP01224 & \footnotesize (minute) & {\bf 0.68}       &  4.4     & {\bf 0.8}& 11  & {\bf 1.7} & 48 \\
 (10.6K) &         & \footnotesize [0.1,1.6] & \footnotesize [3.4,4.9]  & \footnotesize [0.18,1.7] & \footnotesize [9.9,12] & \footnotesize [1.3,1.8] &\footnotesize  [41,55] \\
& & & & & & \\
 HGDP00778 & \footnotesize (hour)   & \bf 0.087       & 0.61        & \bf 0.12      & 1.5           & \bf 0.25        & 6.2 \\
 (78K) &           & \footnotesize [0.03,0.15] & \footnotesize[0.55,0.66] & \footnotesize[0.044,0.12] & \footnotesize[1.3,1.5] & \footnotesize[0.14,0.34] & \footnotesize[3.8,9.4] \\
& & & & & & \\
 HGDP-CEPH & \footnotesize (hour) & \bf 0.9     & 3.7     & \bf  0.92      & 12      & \bf 2.1       & 38 \\
(660K)     && \footnotesize[0.33,1.3] & \footnotesize[3,4.3] &\footnotesize [0.38,1.5] &\footnotesize [11,12] & \footnotesize[1.3,3.0] & \footnotesize[29,45] \\
& & & & & & \\
 1000 Genomes Project & \footnotesize (hour) & \bf 2.8 & (19) & \bf 4.6 & (59) & -- & --\\
 (2.2M) &          & \footnotesize[1.1,4.7] & \footnotesize-- & \footnotesize[1.5,8.3] & \footnotesize-- & \footnotesize-- & \footnotesize-- \\

 \hline
\end{tabular}
\end{table}

\paragraph{Prediction of masked genotypes.} To decide which program options could provide the best estimates, we employed a cross-validation technique based on the imputation of masked genotypes (Wold 1978; Alexander et al. 2011). The cross-validation approach partitions the genotypic matrix entries into a training set and a test set, that are used for estimation and validation sequentially. To build test sets, 5$\%$ of the genotypic matrix entries were tagged as missing values. The masked entries were then predicted using estimates obtained from training sets. Predictions were assessed using a cross-entropy criterion that measured the capability of an algorithm to correctly impute masked genotypes (see Material and Methods). Lower values of the cross-entropy criterion generally indicate better predictive capabilities of an algorithm.  

  Using the cross-entropy criterion, we performed an extensive analysis of {\tt sNMF} program outputs to assess which values of the number of clusters ($K$) and regularization parameter ($\alpha$) could provide the best prediction of masked genotypes (Figure 3). For HGDP data sets with moderate size (panels HGDP00998 and HGDP01224), values of $K$ around 7--8 provided the best predictive results. For larger human data sets, cross-entropy values did not stabilize for $K \leq 10$, indicating that more than ten clusters were necessary to describe population structure. Choices of regularization parameter values greater than 1,000 were generally discarded by the cross-entropy criterion.  For panels of moderate size, the best ancestry estimates were obtained for values around $\alpha = 100$. For panels HGDP00778 and HGDP0927, $\alpha \approx 100-1,000$ provided the best predictive results for most values of $K$. The influence of the regularization parameter was sensitive for the smallest data sets, but for the largest ones a wide range of values led to comparable imputation results. Regardless of the value of the regularization parameter, $K = 5$ clusters led to the best results for The 1000 Genomes Project data set (Figure 3). This last result is in accordance with the criteria used for choosing populations included in The 1000 Genomes Project.  

\centerline{\includegraphics[width=20cm]{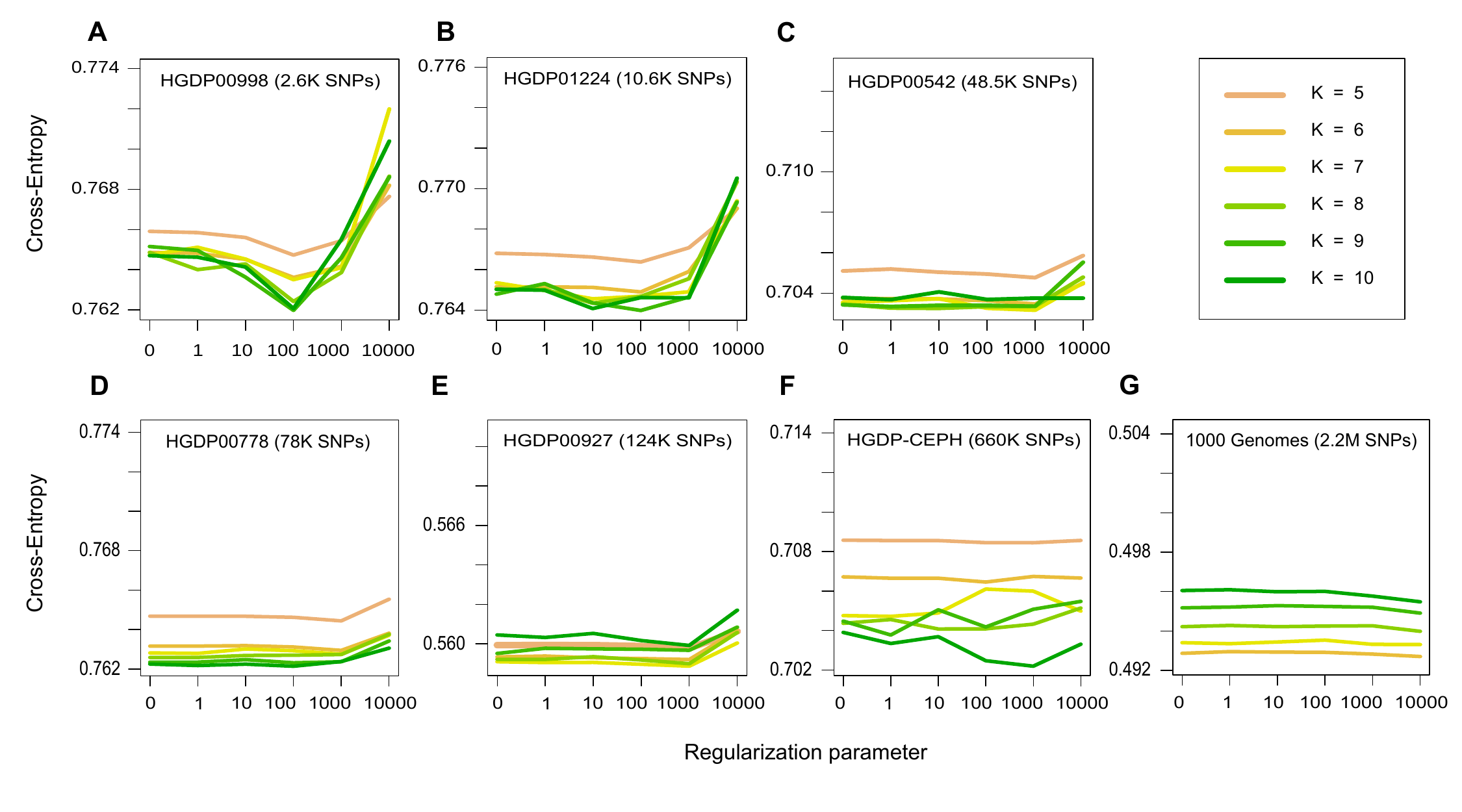}}
{\it 
\noindent{\bf Figure 3.}  {\bf Values of the cross-entropy criterion for {\tt sNMF} algorithms.} Averaged values of the cross-entropy criterion over 5 runs of the {\tt sNMF} program for A-E) 5 Harvard HGDP panel, F) the HGDP-CEPH data, and G) The 1000 Genomes Project data set. The number of clusters ranged from 5 to 10, and  the values of the regularization parameter ranged from 0 to 10,000.
}

 \paragraph{Ancestry estimates.} To compare ancestry estimates obtained from particular runs of {\tt sNMF} and {\tt ADMIXTURE}, we displayed the $Q$-matrix computed by each program for the Harvard HGPD panel HGDP00778 (78K SNPs), the HGDP-CEPH data (660K SNPs), The 1000 Genomes Project phase 1 data (2.2M SNPs) and European populations of {\it A. thaliana} (216K SNPs). Using $K=7$ ancestral populations for the Harvard HGPD panel HGDP00778, the cross-entropy criterion was equal to 0.747 for the {\tt ADMIXTURE} run, and it was equal to 0.762 in the {\tt sNMF} run. The criterion favored {\tt ADMIXTURE} in this case, but the two runs led to very close estimates of the $Q$-matrix ($R^2 =0.99$, Figure S2a). When the programs were applied to the HGDP-CEPH data  ($K= 7$), the cross-entropy criterion was equal to 0.691 for the {\tt ADMIXTURE} run and 0.704 for the {\tt sNMF} run (Figure S2b). This particular {\tt ADMIXTURE} run identified clusters that separated the African hunter-gatherer populations from the other populations, whereas {\tt sNMF} identified a unique cluster in Africa. In the {\tt sNMF} run, Middle East populations were separated from European populations (Figure S2b). The differences between {\tt ADMIXTURE} and {\tt sNMF} results disappeared when additional runs were performed with distinct random seeds.  Using $K = 5$ for The 1000 Genomes Project phase 1 data,  {\tt sNMF} identified clusters that correspond to the main geographic regions of the world, similarly to {\tt ADMIXTURE} (Figure 4, cross-entropy = 0.5010). Substantial levels of European ancestry in African-Americans, Mexican-Americans, Puerto-Ricans and Columbians were inferred by {\tt sNMF} and by the other program. An interesting case was the application of ancestry estimation programs to European populations of {\it A. thaliana}, a selfing plant characterized by high levels of inbreeding (Atwell et al. 2010).  Using $K = 3$, the cross-entropy criterion for {\tt ADMIXTURE} was equal to 0.641 on average, while the average value for {\tt sNMF} was equal to 0.483. The value of the criterion suggests that {\tt sNMF} estimates are more accurate than those obtained from {\tt ADMIXTURE}. The graphical output of the $Q$ matrix displays clinal variation occurring along an east-west gradient separating two clusters.  Northern Swedish accessions were grouped into a separate cluster (Figure S3).  

\centerline{\includegraphics[width=20cm]{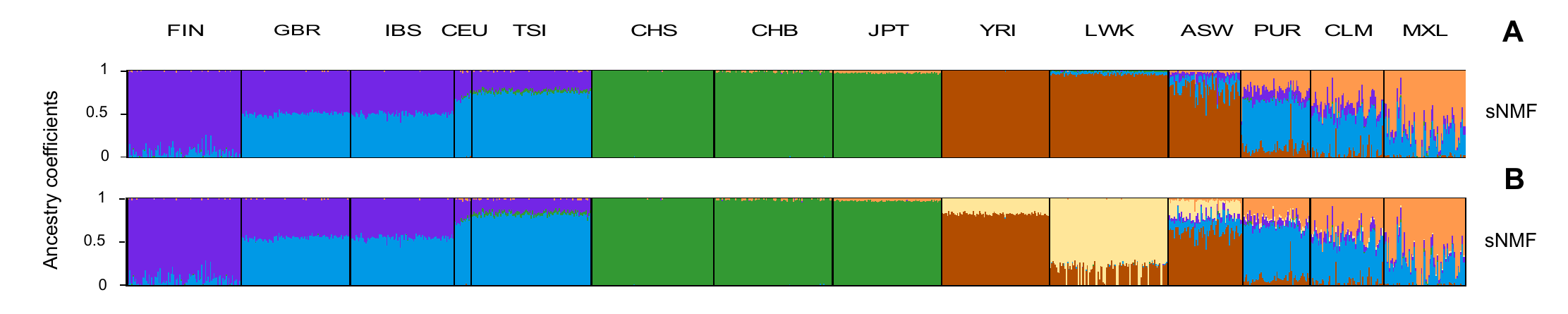}}
{\it 
\noindent{\bf Figure 4.} {\bf Graphical representation of admixture estimates obtained for The 1000 Genomes Project data set.} A) Estimated admixture coefficients using {\tt sNMF} with $K=5$ and $\alpha = 10000$ (cross-entropy = 0.5010). B) Estimated admixture coefficients using {\tt sNMF} with $K=6$ and $\alpha = 10000$ (cross-entropy = 0.5011) (FIN: Finnish, GBR: British, IBS: Spanish, CEU: CEPH Utah residents, TSI: Tuscan, CHS: Southern Han Chinese, CHB: Han Chinese, JPT: Japanese, YRI: Yoruba, LWK: Luhya, ASW: African-American, PUR: Puerto Rican, CLM: Colombian, MXL: Mexican-American).
}

\paragraph{Simulated data analysis.} 

We employed a simulation approach to ascertain the accuracy of {\tt sNMF} estimates and compare them with {\tt ADMIXTURE} estimates. We also assessed the ablility of the cross-entropy criterion to identify the correct value of $K$ when it is known.

In a first series of simulations, we used the 1,000 Genomes Project data to generate genotypes showing various levels of admixture. As our ancestral populations, we chose the CHB, GBR, and YRI samples (The 1,000 Genomes Project 2012), and true $Q$ matrices were created using distinct parameterizations of the Dirichlet distribution (Table 3). Genotypic matrices were simulated according to the binomial model used by {\tt ADMIXTURE}. In this context, {\tt ADMIXTURE} estimates are thus expected to be more accurate than {\tt sNMF} estimates. For the range of parameters explored in the simulations, root mean squared errors comparing the estimated and true value of the $Q$-matrix remained lower than $2\%$ remained for both programs (Table 3). For moderate levels of admixture, differences in statistical errors were lower than $1\%$ regardless of the regularization parameter value, $\alpha$, used in {\tt sNMF}. This result indicates that {\tt sNMF} estimates were
generally accurate, and that relatively small values of $\alpha$ did not influence {\tt sNMF} outputs for data sets of size comparable to those used in simulations.  Root mean squared errors comparing the estimated and true value of the $G$-matrix were slightly lower for {\tt ADMIXTURE} than for {\tt sNMF} (Table 3). This could be explained as we simulated from a binomial model (unrelated individuals) that is assumed by {\tt ADMIXTURE} for the inference of $Q$, and as the number of degrees of freedom in {\tt sNMF}  is twice the number of degrees of freedom in {\tt ADMIXTURE}. 
Regarding the choice of the number of clusters, the cross-entropy criterion was minimal at  the value $K = 3$ for every simulated data sets (Table S2).

\begin{table}
\caption{{\bf Statistical errors for {\tt ADMIXTURE} and {\tt sNMF} on simulated data sets.}}
\begin{center}
\begin{tabular}{lcccccc}
\hline
 & Dir$(1,1,1)$ & Dir$(.5,.5,.5)$ & Dir$(.1,.1,.1)$ & Dir$(.2,.2,.05)$ & Dir$(.2,.2,.5)$ & Dir$(.05,.05,.01)$ \\
\hline
 {~~~~~~\it\small$Q$-matrix}& & & & & & \\
 {\tt ADMIXTURE}         & 0.023 & 0.012 & 0.004 & 0.006 & 0.010 & 0.003 \\
 {\tt sNMF} $\alpha=0$   & 0.020 & 0.011 & 0.007 & 0.007 & 0.013 & 0.009 \\
 {\tt sNMF} $\alpha=100$ & 0.024 & 0.014 & 0.006 & 0.006 & 0.014 & 0.006 \\
\hline
 {~~~~~~\it \small$G$-matrix} & & & & & & \\
 {\tt ADMIXTURE}        & 0.029 & 0.022 & 0.016 & 0.022 & 0.022 & 0.022 \\
 {\tt sNMF} $\alpha=0$   & 0.034 & 0.027 & 0.021 & 0.028 & 0.028 & 0.028 \\
 {\tt sNMF} $\alpha=100$ & 0.034 & 0.027 & 0.021 & 0.028 & 0.028 & 0.028 \\
 \hline
\end{tabular}
\end{center}
Dir: Dirichlet distribution used to simulate "true" admixture coefficients using 3 ancestral populations.
\end{table}

We used a second series of simulated data to evaluate the sensitivity of {\tt ADMIXTURE} and {\tt sNMF} estimates to the presence of related individuals and inbreeding in the sample. Based on empirical data, we used simulation models that mimicked the population structure of European populations of {\it Arabidopsis thaliana}. First we verified that the true value of the number of ancestral populations was correctly recovered by the {\tt sNMF} program using the cross-entropy criterion ($K = 3$). Next we evaluated statistical errors for {\tt ADMIXTURE} and {\tt sNMF} estimates of the $Q$-matrix.   RMSEs remained lower than $4\%$ for both programs. These results showed that the two programs produced accurate estimates of the $Q$-matrix in the presence of inbreeding and missing data (Figure 5). {\tt ADMIXTURE} estimates were robust to the inclusion of moderate levels of inbreeding in the sample, and the {\tt ADMIXTURE} estimates were more accurate than {\tt sNMF} estimates when the value of the inbreeding coefficient was equal to $.25$. For fully inbred lines, {\tt sNMF} produced better estimates than {\tt ADMIXTURE} (Figure 5). The cross-entropy criterion was smaller for {\tt sNMF} than for {\tt ADMIXTURE}, showing that {\tt sNMF} produced better prediction of masked genotypes than {\tt ADMIXTURE} (Figure S4). This result can be explained by a more accurate estimation of genotypic frequencies in {\tt sNMF} than in {\tt ADMIXTURE} in the presence of strong levels of inbreeding.

\centerline{\includegraphics[width=20cm]{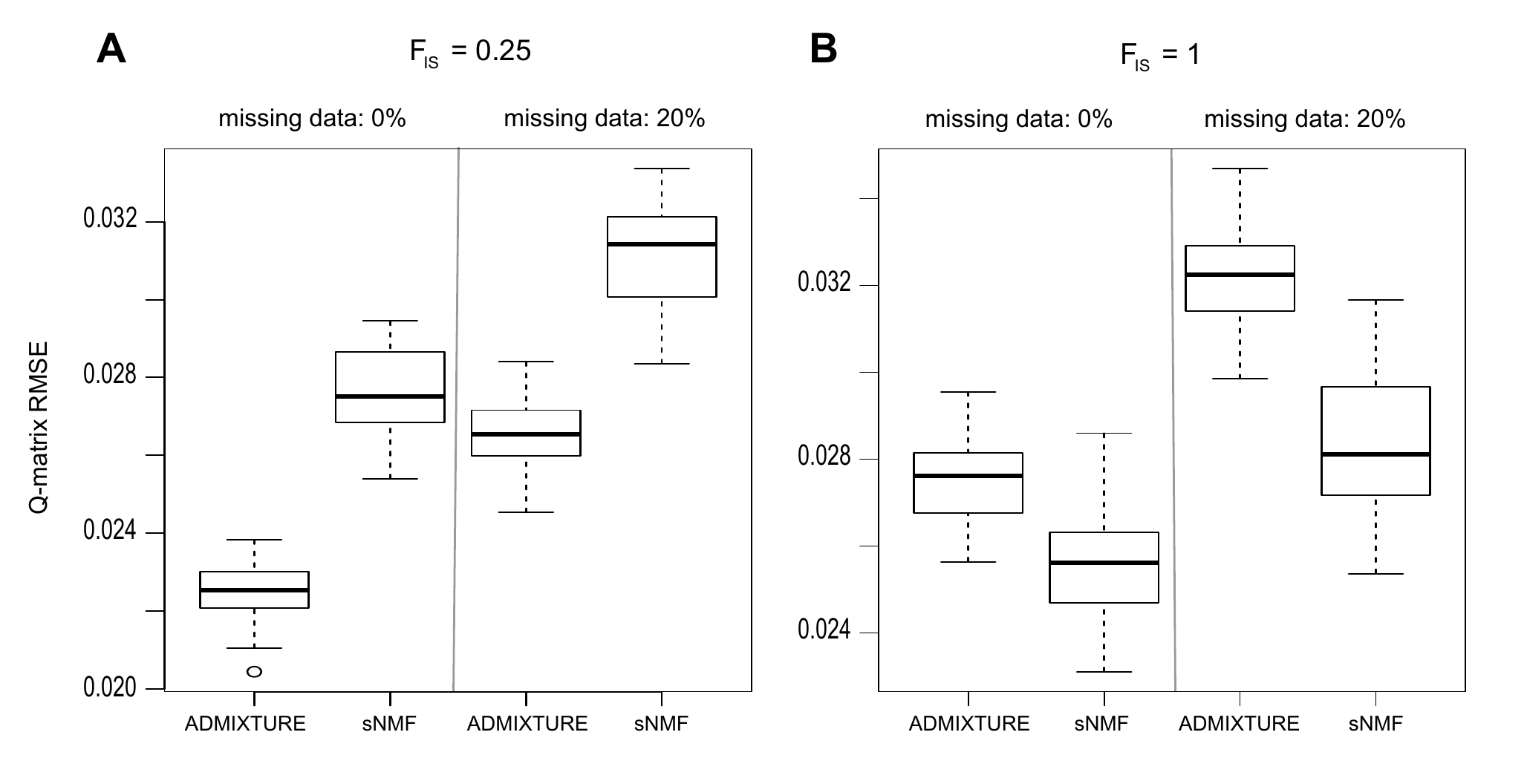}}
{\it 
\noindent{\bf Figure 5.} {\bf Accuracy of {\tt ADMIXTURE} and {\tt sNMF} in the presence of related individuals.} RMSEs between the $Q$-matrix computed by {\tt ADMIXTURE} and by {\tt sNMF} and a known matrix used to generate simulated data. Simulations mimicked the population structure of European populations of {\it Arabidopsis thaliana}.  A)  Moderate and strong levels of inbreeding, $F_{\rm IS} = 25 \%$, B) Strong levels of inbreeding, $ F_{\rm IS} = 100 \%$.
}

\clearpage
\newpage

\section{Discussion}

We applied the computer program {\tt sNMF} to the estimation of individual admixture coefficients using large population genetic data sets for humans and for {\it A. thaliana}, and compared the program performances to those of {\tt ADMIXTURE}. For 6 HGDP data sets, ancestry estimates obtained with {\tt sNMF} and {\tt ADMIXTURE} strongly agreed with each other. In addition, the {\tt sNMF} program was able to analyze The 1000 Genomes project phase 1 data set within a few hours using a standard computer processing unit. Without significant loss of accuracy, {\tt sNMF} computed estimates of admixture proportions within run-times that were about 10 to 30 times faster than those of {\tt ADMIXTURE}. 

The {\tt sNMF} algorithms use theoretical connections between likelihood approaches, PCA and NMF methods (Ding et al. 2008; Engelhardt and Stephens 2010; Lawson et al. 2012; Parry and Wang 2013). 
The statistical literature contains several efficient algorithms to compute NMF estimates, including multiplicative update, projected-gradient, alternate least-square and the active set methods for NMF and sparse NMF problems (Brunet et al. 2004; Berry et al. 2007; Kim and Park 2011).  For population genetic data, we found that the active set method provided the best trade-off between speed and accuracy, and improved significantly over classical NMF implementations (Kim and Park 2011).  
 
To decide which algorithm yielded the best estimates, we introduced a predictive criterion based on the imputation of masked genotypes and the computation of cross-entropy. For HGDP data sets, the cross-entropy criterion discarded large values of the {\tt sNMF} regularization parameter (greater than 1,000). For the largest data sets, a wide range of values of the regularization parameter reached similar predictive values.  For smaller data sets, we found that parsimony (i.e., large values of $\alpha$) could improve estimation of ancestry coefficients. Regarding the significance of the criterion, statistical theory predicts that errors in the cross-entropy criterion are of order $O(1/\sqrt{n_L})$ where $n_L$ is the number of masked genotypes. For Harvard HGDP panels, differences between the {\tt ADMIXTURE} and {\tt sNMF} results were hardly significant, and we considered that both programs provided statistically similar estimates. The examples showed that the cross-entropy criterion could be used to discriminate among program runs regardless of the algorithm used.   

The assumptions underlying {\tt STRUCTURE} and {\tt ADMIXTURE} rely on simplified population genetic hypotheses. More specifically, the assumptions include absence of genetic drift, Hardy-Weinberg and linkage equilibrium in ancestral populations. The factorial coding used by {\tt sNMF} enabled the estimation of homozygote and heterozygote frequencies, and avoided Hardy-Weinberg equilibrium assumptions. Although {\tt ADMIXTURE} analyses  were robust to small departures from Hardy-Weinberg equilibrium in human data, {\tt sNMF} was more appropriate to deal with inbreed populations in {\it A. thaliana}. For the plant data, the values of the cross-entropy criterion indicated better predictive results for {\tt sNMF} than for {\tt ADMIXTURE}. This could be explained as the binomial model of {\tt ADMIXTURE} is not suited to the high levels of inbreeding observed in  {\it A. thaliana} populations (Atwell et al. 2010). The {\tt sNMF} results for European populations of {\it A. thaliana} confirmed previous analyses and PCA predictions based on sequence data (Fran\c cois et al. 2008). 

In conclusion, we found that {\tt sNMF} provided estimates of admixture coefficients that are statistically close to those obtained with {\tt ADMIXTURE}, though likelihood approaches could benefit the analysis of smaller data sets. We found that least-squares NMF approaches were useful for analyzing data where there are obvious departures from standard equilibrium assumptions, for example in populations showing high levels of inbreeding. The overall advantage of the flexible approach taken by {\tt sNMF} was mainly computational, as {\tt sNMF} algorithms were much faster than likelihood algorithms when analyzing very large population genomic data sets.    

\paragraph{Program availability.} The {\tt sNMF} code can be downloaded from the following URL: http://membres-timc.imag.fr/Olivier.Francois/snmf.html.

\clearpage
\newpage

\section{Materials and Methods}

To provide statistical estimates of admixture proportions using multi-locus genotype data sets, we implemented sparse NFM least-squares optimization algorithms in the computer program {\tt sNMF}.

\paragraph{Modeling admixture coefficients.} We considered allelic data for a sample of $n$ multi-locus genotypes at $L$ loci representing single nucleotide polymorphisms (SNPs). The data were stored into a genotypic matrix ($X$) where each entry records the number of derived alleles at locus $\ell$ for individual $i$. For autosomes in a diploid organism, the number of derived alleles at locus $\ell$ is then equal to 0, 1 or 2.  In our algorithm, we used 3 bits of information to encode each  0, 1 or 2 value as an indicator of a heterozygote or a homozygote locus. In other words, the value 0 was encoded as $100$, 1 was encoded as $010$ and 2 as $001$. The use of a factorial coding warrants that the elements on each row of the transformed data matrix sum up to $L$.

Admixture models generally suppose that the genetic data originate from the admixture of $K$ ancestral populations, where $K$ is unknow a priori. Given $K$ populations, the probability that individual $i$ carries $j$ derived alleles at locus $\ell$ can be written as follows
\begin{equation}
p_{i\ell}(j) = \sum_{k = 1}^K q_{ik} g_{k\ell}(j) \,\quad j=  0,1,2, 
\end{equation}
where $q_{ik} $ is the fraction of individual $i$'s genome that originates from the ancestral population $k$, and $g_{k\ell}(j)$ represents the homozygote ($j=0,2$) or the heterozygote ($j=1$) frequency at locus $\ell$ in population $k$. Since it makes no assumption about Hardy-Weinberg equilibrium, the above framework is appropriate to deal with inbreeding and outbreeding in ancestral populations. Using factorial coding, equation (1) writes as 
\begin{equation}
P = Q G  \, ,
\end{equation}
where $P = (p_{i\ell})$ is a $n\times 3L$ matrix, $Q = (q_{ik}) $ is a $n\times K$ matrix, and $G= (g_{k\ell}(j))$ is a $K\times 3L$ matrix. The $Q$ matrix records admixture proportions for each individual in the sample. Though the focus of the above framework is on estimating individual admixture proportions, it can be easily modified to provide ancestry estimates based on allele frequencies in population samples. 
 
\paragraph{Regularized least-squares estimates of ancestry proportions.} 

We approached the inference of admixture proportions using least squares (LS) optimization algorithms (Engelhardt and Stephens 2010). Estimates of the $Q$ and $G$ matrices were obtained after minimizing the following least-squares criterion  
\begin{equation}
{\rm LS}(Q, G) =  \|  X - QG \|^2_{\rm F} \, , 
\end{equation}
where $\| M \|_{\rm F}$ denotes the Frobenius norm of a matrix $M$ (Berry et al. 2007).  Without constraints on $Q$ and $G$, the solutions of the LS problem are given by the singular value decomposition of the matrix $X$, and the resulting matrices $Q$ and $G$ contain the scores and loadings of a PCA. To obtain admixture proportions, $Q$ and $G$ must have non-negative entries such that
\begin{equation} 
\sum_{k=1}^K q_{ik} = 1 \, , \quad \sum_{j=0}^2 g_{k\ell}(j) = 1 \, . 
\end{equation}
With the constraints of equation (4), estimating admixture coefficients and allelic frequencies is equivalent to performing a non-negative matrix factorization of the data matrix, $X$. We introduced a penalty term in equation (3), and we solved a penalized version of the LS optimization problem (3) for $Q$ and $G$ 
\begin{equation}
{\rm SLS}(Q, G) =  \|  X - QG \|^2_{\rm F} + \sqrt{\alpha} \sum_{i =1}^n \| q_{i.}\|^2_1 \, , \quad Q,G \geq 0 \, , 
\end{equation}
with the constraints described in equation (4). In this equation, $\| q_{i.} \|_1$ is the $\ell_1$ norm of $q_{i.}$, and $\alpha$ is a non-negative {\it regularization parameter}.  For $\alpha > 0$, minimizing SLS is equivalent to performing {\it sparse} NMF. Regularized NMF has been previously applied to gene expression data (Kim and Park 2007), and algorithms for sparse NMF were surveyed and compared in (Kim and Park 2011). Large values of $\alpha$ encourage small and large values of admixture proportions, and penalize intermediate levels of ancestry. 
In {\tt sNMF}, estimates of $Q$ and $G$ were computed using the Alternating Non-negativity-constrained Least Squares (ANLS) algorithm with the active set (AS) method (Berry et al. 2007; Kim and Park 2011). We implemented the ALNS-AS method in the computer program {\tt sNMF} using the C programming language. When the number of loci is large compared to the number of individuals, the programming structures used in {\tt sNMF} optimized the time spent in memory access. Several algorithmic methods were also used to speed up the computation of matrix products. While we evaluated run-times using a single computer processor (2.4 GHz 64bits Intel Xeon), a multi-threaded version of the {\tt sNMF} program for multi-processor systems was also developed.

\paragraph{Data sets.}

Admixture inference and run-time analyses were performed on 6 world-wide samples of genomic DNA from  52 populations of the Human Genome Diversity Project -- Centre d'Etude du Polymorphisme Humain (HGDP-CEPH). Five panels were extracted from the Harvard HGDP-CEPH database. These panels were given IDs HGDP00778,  HGDP00542,  HGDP00927, HGDP00998 and HGDP01224, and contained precisely ascertained genotypes of $n = 934$ individuals. The genotypes were specifically designed for population genetic analyses (Patterson et al. 2012). Each marker was ascertained in individuals of Han,Papuan, Yoruba, Karitiana and Mongolian ancestry, and the data matrices included 78,253, 48,531, 124,115, 2,635 and 10,664 SNPs respectively  (Patterson et al. 2012, Table 1). A sample of 1,043 individuals from the HGDP-CEPH Human Genome Diversity Cell Line Panel was also analyzed. The genotypes were generated on Illumina 650K arrays (Li et al. 2008), and the SNP data were filtered to remove low quality SNPs included in the original files. In addition, we used data from The 1000 Genomes Project. The 1000 Genomes Project data contains the genomes of 1,092 individuals from 14 populations, constructed using a combination of low-coverage whole-genome and exome sequencing (phase 1 data, The 1000 Genomes Project Consortium 2012).  The data matrix included  2.2M polymorphic sites across the human genome (Table 1).  

To examine the robustness of {\tt sNMF} to departures from classical population genetic models, additional analyses were performed on a sample of $n = 168$ European accessions from the plant species {\it Arabidopsis thaliana}.  {\it A. thaliana} is a widely distributed self-fertilizing plant known to harbour considerable genetic variation and complex patterns of population structure and relatedness (Atwell et al. 2010).  We analyzed 216,130 SNPs spread across the genome of {\it A. thaliana} (Atwell et al. 2010, Table 1).

\paragraph{Comparisons with {\tt ADMIXTURE}.} The computer program {\tt ADMIXTURE} estimates ancestry coefficients based on the likelihood model implemented in {\tt STRUCTURE}. In {\tt ADMIXTURE}, the assumption of Hardy-Weinberg equilibrium in ancestral populations translates into a binomial model for allele counts at each locus. Considering unrelated individuals, the logarithm of the likelihood can thus be computed as follows
$$
{\cal L} (Q, F) = \sum_i \sum_{\ell} \left(  x_{i \ell}  \log ( \sum_k q_{ik} f_{k\ell}  )    +  ( 1 -x_{i \ell})  \log ( \sum_k q_{ik}(1- f_{k\ell} ) )         \right)
$$
up to an additive constant that does not influence estimation algorithms. In this formula, $Q = (q_{ik}) $ represents the matrix of ancestry coefficients for all individuals, and $F= (f_{k\ell}) $ represents a matrix of allele frequencies for all loci. The $F$ matrix can be converted to a $G$ matrix comparable to the one computed by {\tt sNMF} using the binomial model: $g_{i\ell} (0) = (1 - f_{i\ell})^2 $, $g_{i\ell} (1) = 2 f_{i\ell} (1 - f_{i\ell})$, and  $g_{i\ell} (2) =  f_{i\ell}^2$. {\tt  ADMIXTURE} provides numerical estimates of $Q$ and $F$ that maximize the quantity ${\cal L} (Q, F)$. The local optimization algorithm relies on a block relaxation scheme using sequential quadratic programming for block updates, coupled with a quasi-Newton acceleration of convergence.

A difficulty with optimization algorithms used by {\tt ADMIXTURE} and {\tt sNMF} is that the solutions produced can be dependent on the initial values used for $Q$,  $F$ or $G$.  To enable comparisons with estimates obtained with {\tt ADMIXTURE}, the clusters output by runs of each programs were permuted using {\tt CLUMPP} (Jakobsson et al. 2007).  Differences in ancestry estimates obtained with {\tt ADMIXTURE} ($Q^{\rm ADM}$)  and with {\tt sNMF}  ($Q^{\rm sNMF}$) were assessed by two measures. The first measure was defined as the root mean squared error (RMSE) between the matrices $Q^{\rm ADM}$ and $Q^{\rm sNMF}$  obtained from each program 
$$
{\rm RMSE} =  \left(    \frac{1}{nK} \sum_{i =1}^n \sum_{k=1}^K   ( q_{ik}^{\rm ADM} - q_{ik}^{\rm sNMF})^2   \right)^{1/2} \, .
$$    
Though $G$ matrices could be mainly considered as nuisance parameters for our estimation problem, a similar RMSE criterion was defined for comparing them. The second measure was defined as the squared Pearson correlation coefficient ($R^{2}$) between the matrices $Q^{\rm ADM}$ and $Q^{\rm sNMF}$. When simulations with known $Q$-matrices were analysed, one of the two matrices was replaced by the true $Q$-matrix used to generate the simulated data.

Runs of {\tt ADMIXTURE} and {\tt sNMF} were performed for values of the number of clusters set to $K=5-10$, 15, and 20 for human data sets and set to $K=2-7$ for {\it A. thaliana}. For {\tt sNMF}, the values of the regularization parameter ($\alpha$) ranged between 0 and 10,000 using a $\log_{10}$ scale (5 values).   Each run was replicated five times for a total of 1,410 experiments.  Missing data imputation was initially performed after resampling missing genotypes from the empirical frequency at each locus. The missing values were updated using the predictive probabilities after 20 sweeps of algorithm (see below). Runs of {\tt sNMF} were stopped when the differences between two successive values of the minimization criterion were less in relative value than a tolerance threshold of $\epsilon = 10^{-4}$.

\paragraph{Cross-entropy criterion.} We employed a cross-validation technique  based on imputation of masked genotypes to evaluate the prediction error of ancestry estimation algorithms (Wold 1978, Eastment and Krzanowski 1982).  The procedure partitioned the genotypic matrix entries into a training set and a test set. To build the test set, 5$\%$ of all SNPs were randomly selected, and tagged as missing values. The occurence probabilities for the masked entries were computed using program outputs obtained from training sets with the following formula 
\begin{equation}
p^{\rm pred}_{i\ell}(j) = \sum_{k=1}^K  q_{ik} g_{k\ell}(j)\, , \quad j = 0,1,2 \, . 
\end{equation}
{\tt ADMIXTURE} predicts each masked value by averaging the squares of the deviance residuals for the binomial model (Alexander et al. 2011). Extending the approach employed by {\tt ADMIXTURE} to our non-parametric approach, the predicted values were then compared to the masked values, $x_{i\ell}$,  by averaging the quantity defined as $- \log p^{\rm pred}_{i\ell} (x_{i\ell})$ over all SNPs in the test set.
In statistical terms, our criterion provides an estimate of the following quantity  
\begin{equation}
H(p^{\rm sample}  ,   p^{\rm pred}  ) = -\sum_{j = 0}^2 p^{\rm sample}(j) \log p^{\rm pred}_{i\ell} (j) \, , \quad j = 0,1,2.
\end{equation}
This quantity corresponds to the sum of the Kullback-Leiber divergence between the sample ($p^{\rm sample}$) and predicted ($p^{\rm pred}$) allelic distributions plus the Shannon entropy of the sample distribution. It also corresponds to the cross-entropy between $p^{\rm sample}$ and $p^{\rm pred}$. The number of ancestral gene pools ($K$) and the regularization parameter ($\alpha$) were chosen to minimize the cross-entropy criterion. In general smaller values of the criterion indicate better algorithm outputs and estimates. The standard error of the cross-entropy criterion is of order $1/\sqrt{n_L}$ where $n_L$ is the number of masked genotypes. For data sets including 1,000 individuals genotyped at more than $20, 000$ SNPs, the third digit of the cross-entropy criterion can be significant.  

\paragraph{Simulated data analysis.} 

We adopted a simulation approach to compare RMSEs between the $Q$-matrix computed by {\tt ADMIXTURE} or by {\tt sNMF} and a known matrix used to generate the simulated data.  In addition we assessed whether the correct value of $K$ could be identified by {\tt sNMF} using the cross-entropy criterion.

In a first series of simulations, we used the 1,000 Genomes Project data set to generate artificial data showing various levels of admixture. As our ancestral populations, we chose the CHB, GBR, and YRI samples (The 1,000 Genomes Project 2012). We considered 50,000 SNPs in linkage equilibrium exhibiting no missing genotypes. The allele frequencies observed in our three ancestral populations were used as the true values for the $F$ matrix. The genotypic matrix was constructed according to the binomial model used by {\tt ADMIXTURE}. For 1000 individuals in each simulated data set, a $Q$ matrix was simulated from a Dirichlet probability distribution, and several parameters were explored. Our experiments reproduced the parameters used for evaluating the accuracy of {\tt ADMIXTURE} in a previous study (Alexander et al. 2009).  Runs of {\tt sNMF} were performed for values of the number of clusters set to $K=2-5$ ($\alpha = 0$), and the choice of $K$ was made on the basis on the cross-entropy criterion. For $K = 3$, the values of the regularization parameter ($\alpha$) was varied between 0 and 100. 

Additional data sets were created to mimic the population structure of European populations of {\it Arabidopsis thaliana} using 10,000 SNPs (168 individuals). To defined ancestral frequencies, we used the western European populations grouping samples from the United Kingdom, Belgium and France (23 individuals), central European populations grouping samples from the Czech Republic (24 individuals), and northern European populations grouping samples from Finland and northern Sweden (13 individuals). Both the {\tt ADMIXTURE} and {\tt sNMF} programs grouped those samples within 3 well-separated clusters exhibiting low levels of admixture with other plant populations. The empirical frequencies computed from the 3 populations were considered as the true frequencies in our simulation procedure considering a generative model with $K = 3$ ancestral populations. From empirical frequencies, we computed genotypic frequencies, $f_{k\ell}$, using 2 distinct values of population inbreeding coefficient, $F_{\rm IS} = 25 \%$ and $ F_{\rm IS} = 100 \%$, that corresponded to moderate and strong levels of inbreeding. For 168 individuals, 10,000 genotypes were simulated using the sampling equation $p(x_{i\ell} = j) = \sum_k q_{ik} g_{k\ell}(j)$, where $q_{ik}$ corresponds to the $Q$-matrix computed from the full empirical data set (216K SNPs). In addition,  simulated data sets were generated with or without missing data (0 or 20 $\%$). Fifty replicates were created for each value of the inbreeding coefficient and for each value of the ratio of missing data.

\section*{References}
\begin{description}
\item Alexander DH, Novembre J, Lange K. 2009. Fast model-based estimation of ancestry in unrelated individuals. \textit{Genome Res} \textbf{19}: 1655--1664.

\item Alexander DH, Lange K. 2011. Enhancements to the admixture algorithm for individual ancestry estimation. \textit{BMC Bioinformatics}  \textbf{12}: 246.

\item Atwell S, Huang YS, Vilhj\'{a}lmsson BJ, Willems G, Horton M, Li Y, Meng D, Platt A, Tarone AM, Hu TT, et al.	2010. Genome-wide association study of 107 phenotypes in {\it Arabidopsis thaliana} inbred lines. \textit{Nature} \textbf{465}: 627--631.



\item Berry MW, Browne M, Langville AN, Pauca VP, Plemmons RJ. 2007. Algorithms and applications for approximate nonnegative matrix factorization. \textit{Comput Stat Data An} \textbf{52}: 155--173.


\item Brunet JP, Tamayo P, Golub TR, Mesirov JP. 2004. Metagenes and molecular pattern discovery using matrix factorization. \textit{P Natl A Sci} \textbf{101}: 4164--4169.


\item Cavalli-Sforza LL, Bodmer WF. 1971. \textit{The genetics of human populations.} Dover Publications, Mineola, NY.









\item Ding C, Li T, Peng W. 2008. On the equivalence between non-negative matrix factorization and probabilistic latent semantic indexing. \textit{Comput Stat Data An} \textbf{52}: 3913--3927.


\item Eastment HT, Krzanowski WJ. 1982. Cross-validatory choice of the number of components from a principal component analysis. \textit{Technometrics} \textbf{24}: 73--77.

\item Engelhardt BE, Stephens M. 2010. Analysis of population structure: a unifying framework and novel methods based on sparse factor analysis. \textit{PLoS Genet} \textbf{6}: 12.



\item Fran\c{c}ois O, Blum MGB, Jakobsson M, Rosenberg NA. 2008. Demographic history of European populations of {\it Arabidopsis thaliana}. \textit{PLoS Genet} \textbf{4}: e1000075.


\item Fran\c{c}ois O, Currat M, Ray N, Han E, Excoffier L, Novembre J. 2010. Principal component analysis under population genetic models of range expansion and admixture. \textit{Mol Biol Evol} \textbf{27}: 1257--1268.


\item Frichot E, Schoville SD, Bouchard G, Fran\c{c}ois O. 2012. Correcting principal component maps for effects of spatial autocorrelation in population genetic data. \textit{Front Genet} \textbf{3}: 254.

\item Frichot E, Schoville SD, Bouchard G, Fran\c{c}ois O. 2013.  Testing for associations between loci and environmental gradients using latent factor mixed models. \textit{Mol Biol Evol}  \textbf{30}: 1687--1699.







\item Kim H, Park H. 2007. Sparse non-negative matrix factorizations via alternating non-negativity-constrained least squares for microarray data analysis. \textit{Bioinformatics} \textbf{23}: 1495--1502.


\item Kim J, Park H. 2011. Fast Nonnegative Matrix Factorization: An active-set-like method and comparisons. \textit{SIAM J Sci Comput} \textbf{33}: 3261--3281.





\item Jakobsson M, Rosenberg NA. 2007. CLUMPP: a cluster matching and permutation program for dealing with label switching and multimodality in analysis of population structure. \textit{Bioinformatics} \textbf{23}: 1801--1806.

\item Lawson DJ, Falush D. 2012. Population identification using genetic data. \textit{Annu Rev Genom Hum G} \textbf{13}: 337--361.

\item Lawson DJ, Hellenthal G, Myers S, and Falush D. 2012. Inference of population structure using dense haplotype data. \textit{PLoS Genet} \textbf{8}: e1002453.

\item Lee DD, Seung HS. 1999. Learning the parts of objects by non-negative matrix factorization. \textit{Nature} \textbf{401(6755)}: 788--791.

\item Li JZ, Absher DM, Tang H, Southwick AM, Casto AM, Ramachandran S, Cann HM, Barsh GS, Feldman M, Cavalli-Sforza LL, Myers RM. 2008. Worldwide human relationships inferred from genome-wide patterns of variation. \textit{Science} \textbf{319}: 1100--1104.





\item Marchini J, Cardon LR, Phillips MS, Donnelly P. 2004. The effects of human population structure on large genetic association studies. \textit{Nat Genet} \textbf{36}: 512--517.


\item McVean G. 2009. A genealogical interpretation of principal components analysis. \textit{PLoS Genet} \textbf{5}: 10.

\item Novembre J, Stephens M. 2008. Interpreting principal component analyses of spatial population genetic variation. \textit{Nat Genet} \textbf{40}: 646--649.

\item Parry RM, Wang MD. 2013. A fast least-squares algorithm for population inference. \textit{BMC bioinformatics} \textbf{14}: 28.

\item Patterson N, Price A, Reich D. 2006. Population structure and eigenanalysis. \textit{PLoS Genet} \textbf{2}: e190.

\item Patterson NJ, Moorjani P, Luo Y, Mallick S, Rohland N, Zhan Y, Genschoreck T, Webster T, Reich D. 2012. Ancient admixture in human history. \textit{Genetics} \textbf{192}: 1065--1093.


\item Price, AL, Patterson, NJ, Plenge RM, Weinblatt ME, Shadick NA, Reich D. 2006. Principal components analysis corrects for stratification in genome-wide association studies. \textit{Nat genet} \textbf{38}: 904--909.

\item Pritchard JK, Stephens M, Donnelly P. 2000. Inference of population structure using multilocus genotype data. \textit{Genetics} \textbf{155}: 945--959.

\item Pritchard JK, Stephens M, Rosenberg NA, Donnelly P. 2000. Association mapping in structured populations. \textit{Am J Hum Genet} \textbf{67}: 170.

\item Roberts DF, Hiorns RW. 1965. Methods of analysis of the genetic composition of a hybrid population. \textit{Hum Biol} \textbf{37}: 38--43.

\item Tang H, Peng J, Wang P, Risch N. 2005. Estimation of individual admixture: Analytical and study design considerations. \textit{Genet Epidemiol} \textbf{28}: 289--301. 
\item The 1000 Genomes Project Consortium. 2012. An integrated map of genetic variation from 1,092 human genomes. \textit{Nature} \textbf{491}: 56--65.

\item Wold S. 1978. Cross-validatory estimation of the number of components in factor and principal components models. \textit{Technometrics} \textbf{20}: 397--405.



\end{description}

\clearpage
\newpage

\section*{Supplementary Materials}

\clearpage
\newpage

\begin{table}
\noindent{{\bf Table S1:} {\bf Root mean square error (RMSE) for 6 HGDP data sets as a function of the
regularization parameter.}}
\begin{center}
\begin{tabular}{cccccc}

 {\bf $\alpha$} & 0 & 1& 10 & 100 & 1000 \\

\hline

\multirow{2}{*}{\bf RMSE} & 0.046 & 0.044 & 0.041 & 0.041 & 0.055 \\

  & [0.035,0.064] & [0.035,0.057]& [0.035,0.052] & [0.031,0.061] & [0.033, 0.095]\\

\end{tabular}
\end{center}
\end{table}

\clearpage
\newpage

\begin{table}
\noindent{{\bf Table S2:} {\bf Choice of $K$ for {\tt sNMF} using the cross-entropy criterion (simulated data).}}
\begin{center}
\begin{tabular}{lcccccc}
 &  Dir$(1,1,1)$ & Dir$(.5,.5,.5)$ & Dir$(.1,.1,.1)$ & Dir$(.2,.2,.05)$ & Dir$(.2,.2,.5)$ & Dir$(.05,.05,.01)$ \\
\hline
 $K=2$ & 0.713 & 0.703 & 0.682 & 0.662 & 0.706 & 0.645 \\
 $K=3$ & {\bf 0.707} & {\bf 0.691} & {\bf 0.660} & {\bf 0.642} & {\bf 0.697} & {\bf 0.624} \\
 $K=4$ & 0.708 & 0.692 & 0.661 & 0.644 & 0.699 & 0.626 \\
 $K=5$ & 0.710 & 0.694 & 0.663 & 0.645 & 0.700 & 0.628 \\
\end{tabular}
\end{center}
Dir: Dirichlet distribution used to simulate "true" admixture coefficients using 3 ancestral populations.
\end{table}

\clearpage
\newpage

\centerline{\includegraphics[width=15cm]{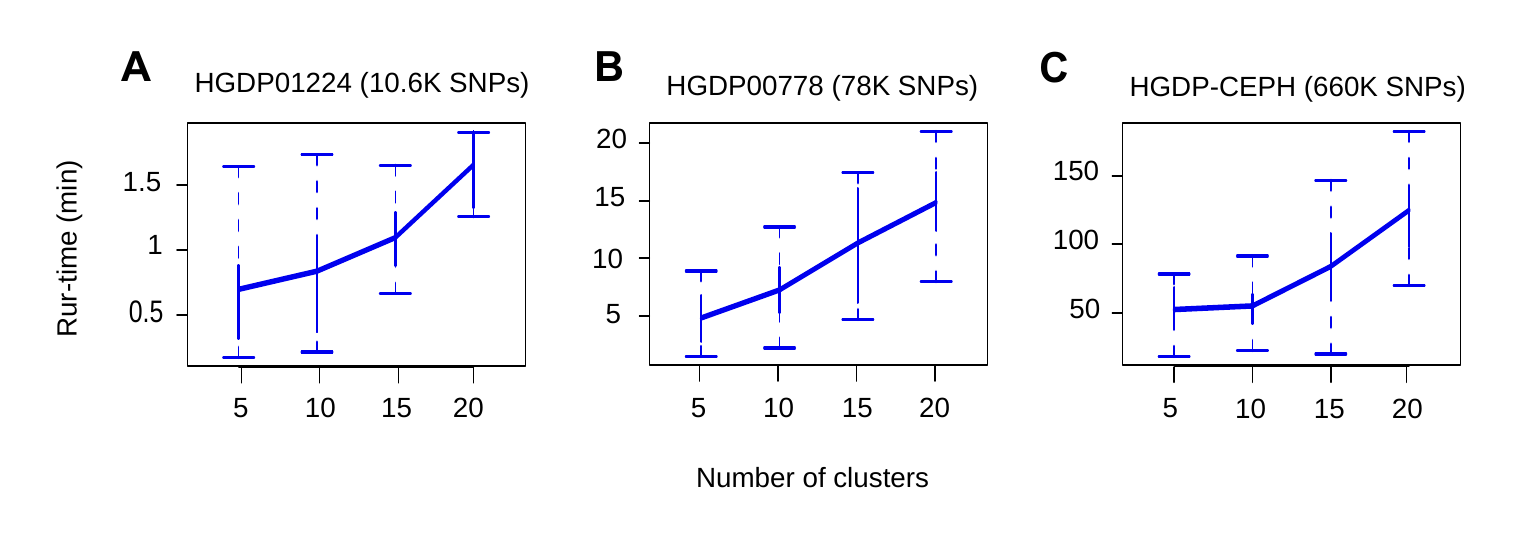}}
\noindent{\bf Figure S1.} {\bf Run-times for {\tt sNMF}.} Time is expressed in unit of minutes. A) Run-time analysis for Harvard HGDP panel 01224 (10.6K SNPs). B) Run-time analysis for Harvard HGDP panel 00778 (78K SNPs). C) Run-time analysis for the HGDP-CEPH data (660K SNPs).

\clearpage
\newpage

\noindent
\centerline{
\includegraphics[width=20cm]{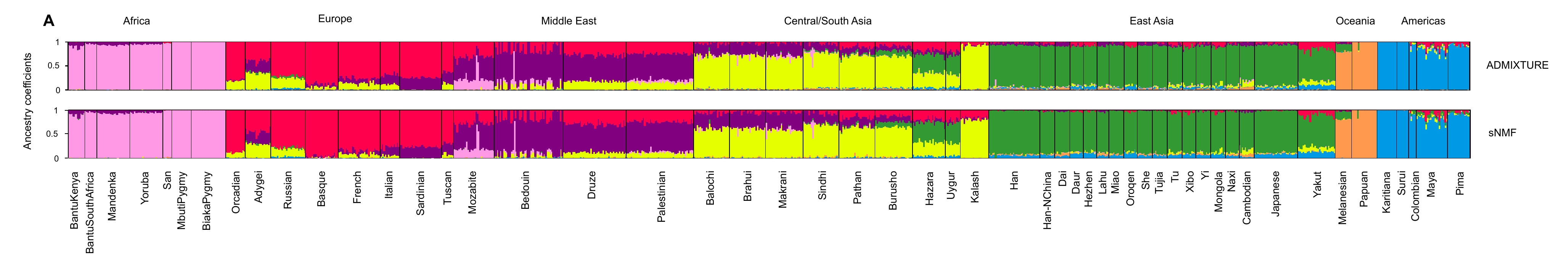}
}
\\
\noindent
\centerline{
\includegraphics[width=20cm]{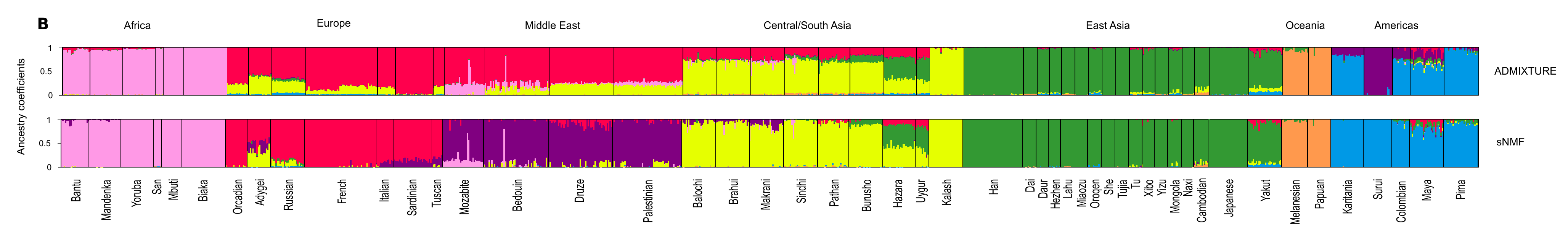}
}
\\
\noindent{\bf Figure S2.} {\bf Graphical representation of admixture estimates obtained for two human data sets  ($K=7$).} A) HGDP00778 panel (78K SNPs). Estimated admixture coefficients using {\tt ADMIXTURE} (top, cross-entropy = 0.747) and {\tt sNMF} (bottom, cross-entropy = 0.762 and $\alpha=1,000$). B) HGDP-CEPH data set. Estimated admixture coefficients using {\tt ADMIXTURE} (top, cross-entropy = 0.691) and {\tt sNMF} (bottom, cross-entropy = 0.704 and $\alpha=1,000$).

\clearpage
\newpage

\centerline{\includegraphics[width=18cm]{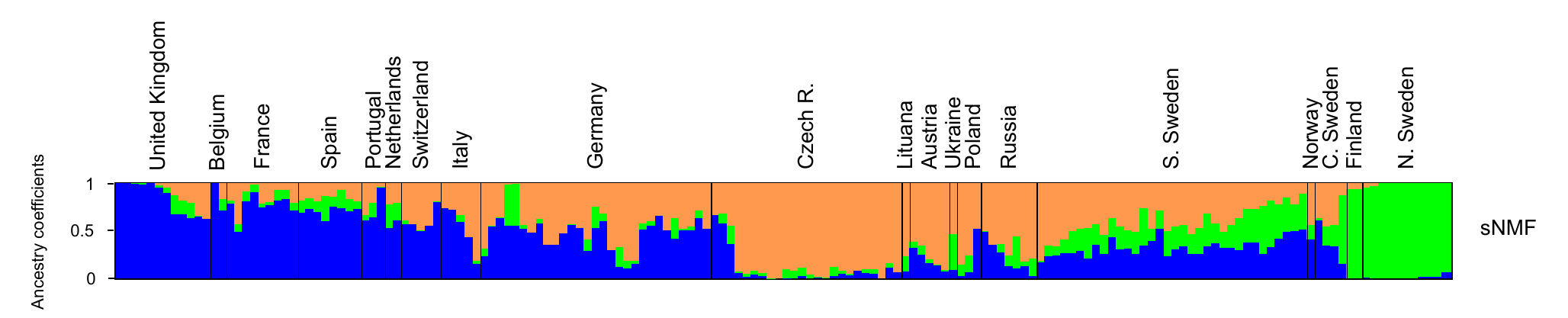}}
\noindent{\bf Figure S3.} {\bf Graphical representation of admixture estimates for European populations of {\it A. thaliana}.} Estimated admixture coefficients using {\tt sNMF} using $K=3$ and $\alpha=100$ (cross-entropy = 0.483).

\clearpage
\newpage

\centerline{\includegraphics[width=18cm]{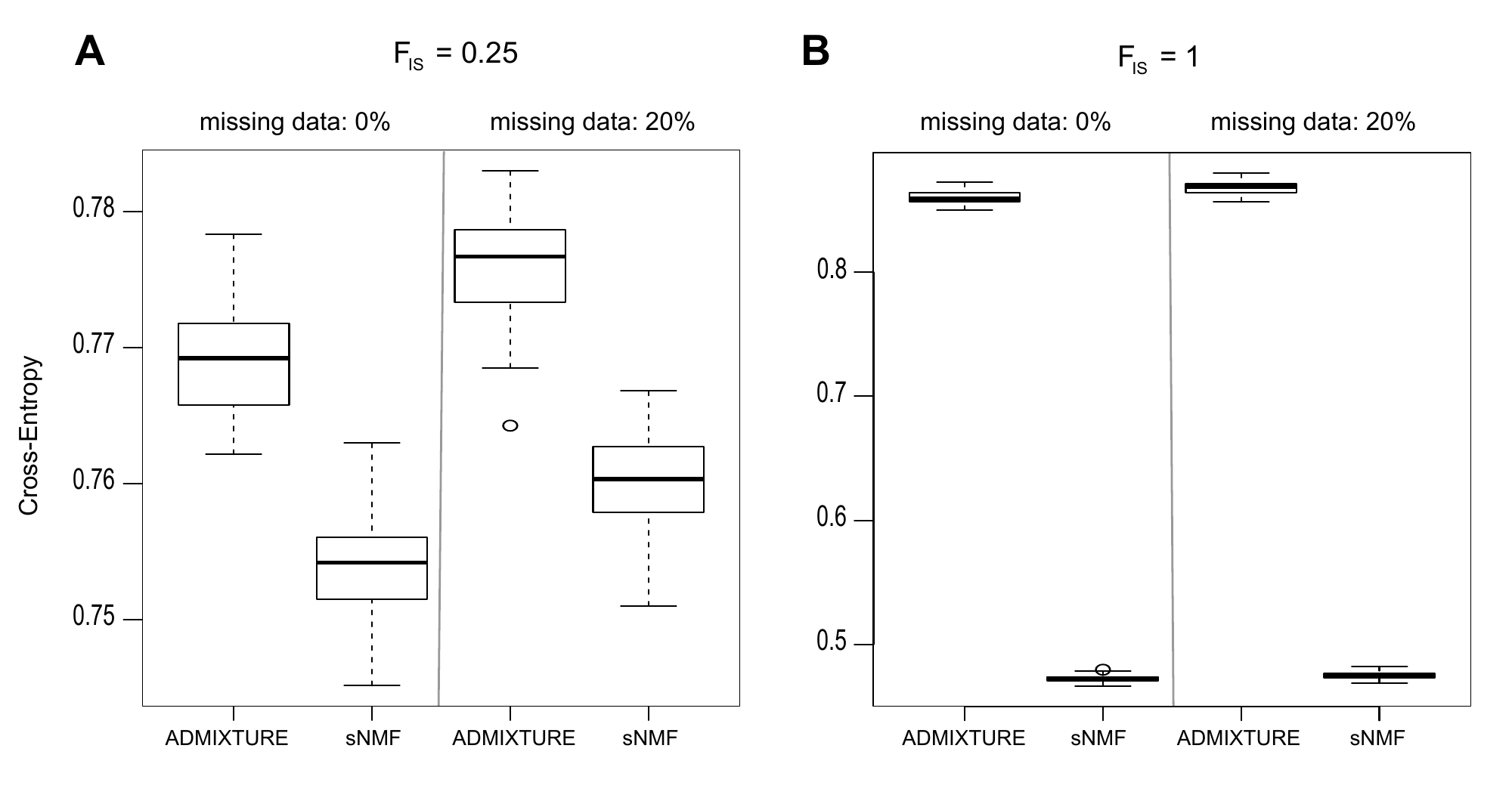}}
\noindent{\bf Figure S4.} {\bf Accuracy of {\tt ADMIXTURE} and {\tt sNMF} in the presence of related individuals.}
Cross-entropy criterion for {\tt ADMIXTURE} and for {\tt sNMF}.
Simulations mimicked the population structure of European populations of {\it Arabidopsis thaliana}.  A)  Moderate and strong levels of inbreeding, $F_{\rm IS} = 25 \%$, B) Strong levels of inbreeding, $ F_{\rm IS} = 100 \%$.

\end{document}